\documentclass[aps,prd,twocolumn,superscriptaddress,showpacs,floatfix]{revtex4-1} 
\usepackage{graphicx}
\usepackage{dcolumn}   
\usepackage{bm}        
\usepackage{amssymb}   
\usepackage{color}
\usepackage{palatino}
\usepackage{ifluatex}
\usepackage[utf8]{\ifluatex lua\fi inputenc}
\usepackage[T1]{fontenc}
\usepackage{newtxtext,newtxmath}
\usepackage{epstopdf}
\usepackage{slashed}
\usepackage{lineno}
\usepackage[FIGTOPCAP]{subfigure}
\usepackage{stackengine}

\begin{document}
\title{Double-strangeness production in $\Lambda p\to K^+X$ reaction}
\author{Jung Keun Ahn}
\affiliation{Department of Physics, Korea University, Seoul 02841, 
Republic of Korea}
\author{Seung-il Nam\footnote{E-mail: {\tt sinam@pknu.ac.kr}}}
\affiliation{Department of Physics and Institute for Radiation Science \& Technology (IRST), 
Pukyong National University, Busan 48513, Republic of Korea}
\affiliation{Asia Pacific Center for Theoretical Physics (APCTP), Pohang 37673, 
Republic of Korea}\date{\today}

\date{\today}

\begin{abstract}
We investigate $S=-2$ production from 
the $\Lambda p\to K^+X$ reactions 
within the effective Lagrangian approach. 
The $\Lambda p\to K^+\Lambda\Lambda$ and 
$\Lambda p\to K^+\Xi^-p$ reactions are considered to find 
the lightest $S=-2$ system, which is
$H$-dibaryon. We assume that 
the $H(2250)\to\Lambda\Lambda$, and $H(2270)\to\Xi^-p$
decays with the intrinsic decay width of 1 MeV.
According to our calculations, the total cross-sections 
for $\Lambda p\to K^+\Lambda\Lambda$ and $\Lambda p\to K^+\Xi^-p$ reactions
were found to be of the order of a few
$\mu$b in the $\Lambda$ beam momentum range of 
up to 5 GeV$/c$. Furthermore,  
the direct access of information regarding the interference patterns
between the $H$-dibaryon and non-resonant contributions was
demonstrated.
\end{abstract}
\maketitle


\section{Introduction}

Double-strangeness baryon systems involve an $H$-dibaryon, 
double hypernuclei, and possibly the inner core of neutron stars \cite{Kahana}. 
An observation of several double hypernuclei 
reveals that the $\Lambda\Lambda$ interaction is weakly attractive. 
However, the $\Xi^-N$ interaction was only studied 
in heavy-ion collisions, which indicates a strong, 
attractive interaction \cite{alice}. 
Recently, the $\Xi N$--$\Lambda\Lambda$ coupling 
was determined to be weak based on an initial observation 
of a Coulomb-assisted bound state for the
$\Xi^-$--$^{14}$N system \cite{shuhei}, which was predicted to exist 
considering the evidence for a deeply-bound $\Xi^-$--$^{14}$N state
reported in a hybrid emulsion experiment at KEK-PS \cite{nakazawa}.  
While strangeness
$S=-2$ baryon-baryon interactions provide critical information on
exploring the smallest object ($H$-dibaryon), 
and the largest (the inner core of 
neutron stars), the experimental data is limited. 

The lightest $S=-2$ system is 
the $H$-dibaryon, which can be decomposed into a compact 6-quark state, 
and two baryon states involving $\Lambda\Lambda$, $\Xi N$, 
and $\Sigma\Sigma$ components. 
The mass range of the $H$-dibaryon is strongly connected with 
the existence of double $\Lambda$ hypernuclei. 
Several double $\Lambda$ hypernuclei have been reported: 
${}^{~~6}_{\Lambda\Lambda}$He \cite{takahashi}, 
${}^{~10}_{\Lambda\Lambda}$Be \cite{ekawa}, 
and ${}^{~13}_{\Lambda\Lambda}$B \cite{aoki}. 
Because the $\Lambda\Lambda\to H$ decay
was not observed in the aforementioned studies, the $H$ must be
heavier than $m_H>2m_{\Lambda}+B_{\Lambda\Lambda}\approx 2.22$ GeV$/c^2$.

Recently, the HAL QCD collaboration has indicated 
that the $\Lambda\Lambda$($^{1}$S$_0$) interaction is not sufficiently 
attractive to generate a bound or resonant state 
close to the $\Lambda\Lambda$ threshold, whereas 
the $\Xi N$($^{1}$S$_0$) phase shift increases sharply 
just above the $\Xi N$ threshold \cite{sasaki}.
Experimental confirmation of the $H$-dibaryon would be 
a significant accomplishment for a better understanding of hyperon interactions.

Enhanced $\Lambda\Lambda$ production close to the $\Lambda\Lambda$ threshold
was reported in $^{12}$C$(K^-,K^+)$ reactions at $p_{K^-}=1.65$ GeV$/c$
\cite{e224,e522}. This threshold enhancement may provide insight for 
the possible existence of an $H$-dibaryon near the $\Lambda\Lambda$ or $\Xi^-p$
thresholds. A high-statistics experimental reconfirmation should be awaited 
until the dedicated $H$-dibaryon search experiment E42\cite{e42} is performed 
using a high-intensity $K^-$ beam at J-PARC. 
  
The simplest method for producing the $H$-dibaryon is to employ 
the double-strangeness and double-charge exchange 
$(K^-,K^+)$ reaction on a light nuclear target 
to retain two units of strangeness in a $^{12}$C nucleus,
similar to the J-PARC E42 with a diamond target at $p_{K^-}=1.8$ GeV$/c$. Furthermore, the $H$-dibaryon is also available in other reactions, such as 
$pp$, $pA$, $\gamma A$, and $\pi A$, most of which involve nuclear
targets that contain at least two nucleons coupled to the $H$-dibaryon production; therefore, 
the overlap of wavefunctions for hyperons and intranuclear nucleons should be 
considered. A cross-section measurement for the $\Lambda\Lambda$
production was reported to be $6.7\pm 1.5$ mb 
in a $\overline{p}$Ta reaction at 4 GeV$/c$ \cite{miyano}.
Heavy-ion reactions can be used to produce $\Lambda$ and $\Xi^-$ 
hyperons copiously so that the coalescence of 
two of these particles into the $H$-dibaryon may be observed. 
However, the H-dibaryon should be observed 
in a high-multiplicity environment for high-energy heavy-ion collisions. 
 
Because the $H$-dibaryon can be formed directly via $\Xi N$ and 
$\Lambda\Lambda$ interactions, the production reaction 
involving the minimum number of vertices is the
$\Xi^-p\to H$ reaction with a proton target. However, 
in this case, the mass range of the $H$-dibaryon is accessible only above 
the $\Xi^-p$ mass threshold. Because a $\Lambda\Lambda$ scattering experiment
is unavailable, the second-best choice is a $\Lambda p\to HK^+$ reaction via
a strangeness-exchange process, 
with which the $H$-dibaryon can be observed 
in the mass range below the $\Lambda\Lambda$ threshold to a higher mass region.     

A $\Lambda$ beam is available via photoproduction and $\pi^-$-induced reactions
by tagging $K^+$, $K^0$, or $K(892)^\ast$ in the final state. For example, 
the $\pi^-$-induced reactions can either be a 
$\pi^-p\to K^0\Lambda$ or $\pi^-p\to K(892)^\ast\Lambda$ reaction. 
As the detection of a $K^0_S\to 2\pi$ decay triggers  
the production of both $S=+1$ $K^0$ and $S=-1$ $\overline{K}^0$ 
with nearly equal
probability, the production of $\Lambda$ particles cannot be uniquely tagged. 
Therefore, the $\pi^-p\to K(892)^\ast\Lambda$ reaction is selected as a primary
reaction for the $\Lambda p$ elastic scattering measurement 
using an 8 GeV$/c$ $\pi^-$ beam at J-PARC \cite{honda}. 
In this case, the $\Lambda$ beam is available in the momentum
ranging from 0.2 to 2.0 GeV$/c$, 
and it is unavailable for double-strangeness 
production above the threshold $\Lambda$ momentum of 2.6 GeV$/c$. 

However, the $\gamma p\to K^+\Lambda$ reaction 
facilitates the production of a high-momentum $\Lambda$ with 
$\Lambda$ polarization in the photon beam energy region above 2.5 GeV. 
The measurement of $\Lambda p\to K^+\Lambda\Lambda$ and $\Lambda p\to K^+\Xi^-p$
is viable with the CLAS data \cite{clas} and 
the upcoming LEPS2 data \cite{leps2}. 
This $(\Lambda,K^+)$ reaction measurement leads to a decisive conclusion
regarding the existence of the H-dibaryon near the $\Lambda\Lambda$ 
and $\Xi^-p$ thresholds.   
Moreover, possible interference effects among 
the $K^+\Lambda\Lambda$ and $K^+\Xi^-p$ channels are noteworthy. 
   
In this study, numerical calculation results 
for the $\Lambda p\to K^+\Lambda\Lambda$ and 
$\Lambda p\to K^+\Xi^-p$ reactions within the effective Lagrangian approach
have been reported. 
We calculate the Dalitz plot densities
$(d^2\sigma/dM_{\Lambda\Lambda}dM_{\Lambda K^+})$ for 
the $\Lambda p\to K^+\Lambda\Lambda$
reaction and $(d^2\sigma/dM_{\Xi^-p}dM_{\Xi^-K^+})$ 
for the $\Lambda p\to K^+\Xi^-p$ reaction.
The $H$-dibaryon states are assumed to
appear at 2.25 GeV$/c^2$ and 2.27 GeV$/c^2$ in the $\Lambda\Lambda$ and 
$\Xi^-p$ channels, respectively. The intrinsic width of the $H$-dibaryon
was chosen to be 1 MeV.
Based on calculations, the total cross-sections 
for the $\Lambda p\to K^+\Lambda\Lambda$ 
and $\Lambda p\to K^+\Xi^-p$ reactions were determined to be within 
the order of a few $\mu$b in the $\Lambda$ beam momentum of 
up to 5 GeV$/c$.
Furthermore, we demonstrated that 
information regarding the interference patterns
between the $H$-dibaryon and non-resonant contributions
can be directly accessed.

\section{Theoretical framework}

\begin{figure}[!hpbt]
\centering
\includegraphics[width=9cm]{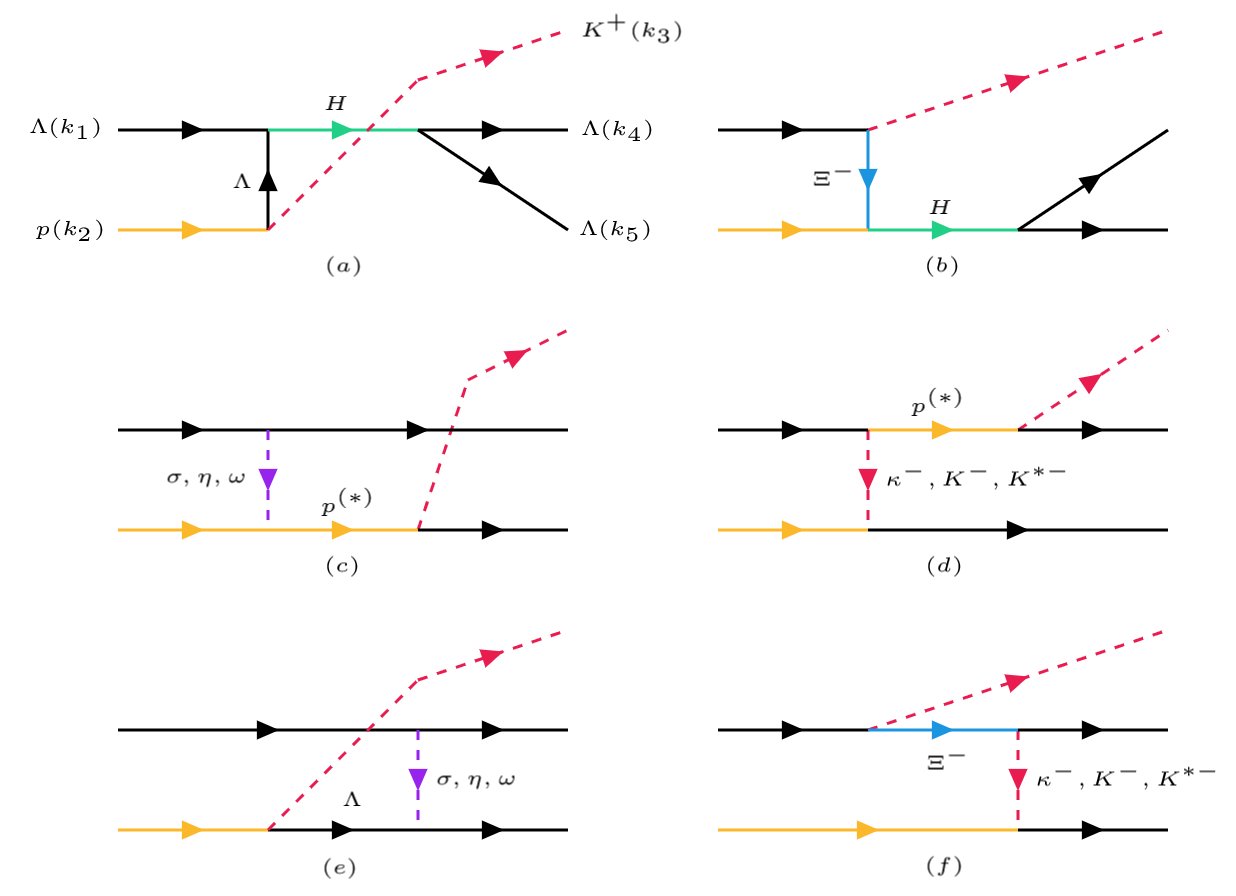}
\caption{(Color online) Relevant Feynman diagrams 
for the present reaction processes of 
$\Lambda p \to K^+\Lambda\Lambda$ at the tree level. 
Diagrams $(a)$ and $(b)$ indicate the $H$-dibaryon-pole contributions 
with the $\Lambda$ and $\Xi$ exchanges. 
Diagrams ($c$--$f$) denote the various baryon-pole contributions 
with the strange 
and nonstrange meson exchanges in the $t$-channel. }\label{fig:fig1}
\end{figure}

\begin{figure}[!thpb]
\centering
\includegraphics[width=9cm]{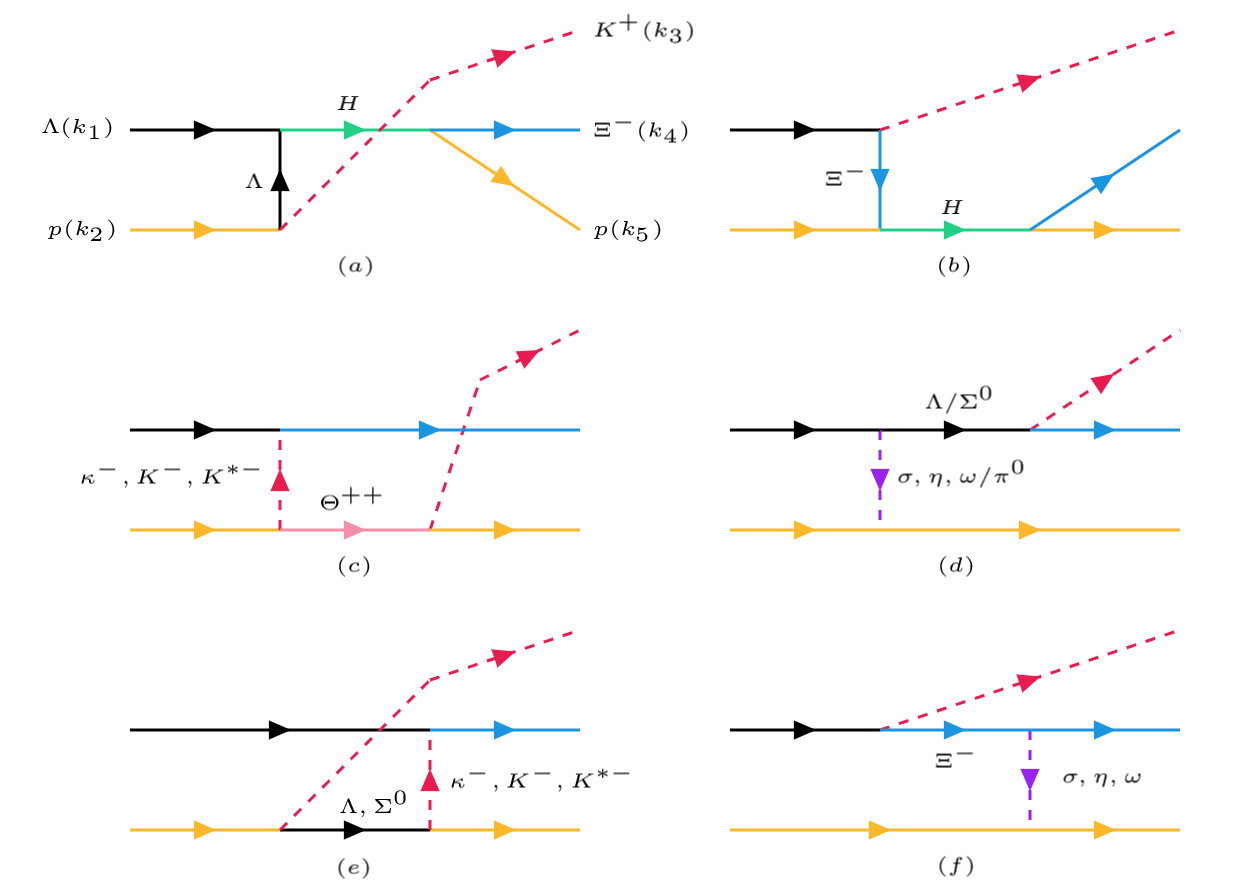}
\caption{(Color online) Corresponding Feynman diagrams 
for the reaction processes of 
$\Lambda p\to K^+\Xi^-p$ at the tree level. 
Diagrams $(a)$ and $(b)$ indicate 
the $H$-dibaryon-pole contributions with the $\Lambda$ and $\Xi$ exchanges. 
Diagrams ($c$--$f$) denote 
the various baryon-pole contributions with the strange 
and non-strange meson exchanges in the $t$-channel. }\label{fig:fig2}\end{figure}

In this section, we introduce the theoretical framework to investigate
the $\Lambda p\to K^+X$ reactions within the effective Lagrangian approach.
We consider the $\Lambda p\to K^+\Lambda\Lambda$ 
and $\Lambda p\to K^+\Xi^-p$ reactions. 
The relevant Feynman diagrams for the reactions are illustrated 
in Figs.~\ref{fig:fig1} and \ref{fig:fig2}.
Diagrams ($a$) and ($b$) indicate an $H$-dibaryon-pole with 
$\Lambda$ and $\Xi$ exchanges, respectively. 
The other diagrams ($c$--$f$) 
denote various baryon-pole contributions with $t$-channel meson exchanges. 
The effective Lagrangians for the Yukawa vertices are defined as follows:
\label{eq:LSG}
\begin{eqnarray}
\mathcal{L}_{BBH}&=&-g_{BBH}B^\dagger B^\dagger H+\mathrm{h.c.}, \nonumber \\
\mathcal{L}_{BBH}&=&g_{BBBB}B^\dagger B^\dagger \Gamma BB+\mathrm{h.c.}, \nonumber \\
\mathcal{L}_{SBB}&=&-ig_{SBB}\bar{B}S B+\mathrm{h.c.}, \nonumber \\
\mathcal{L}_{PBB}&=&-ig_{MBB}\bar{B}\gamma_5 P B+\mathrm{h.c.}, \nonumber \\
\mathcal{L}_{PBB'}&=&-ig_{MBB'}\bar{B}'P B+\mathrm{h.c.}, \nonumber \\
\mathcal{L}_{VBB}&=&g_{VBB}\bar{B}\rlap{/}{V} B+\mathrm{h.c.}, 
\end{eqnarray}
%
where $B$, $B'$, $H$, $S$, $P$, and $V$ denote the fields 
for the $1/2^+$ baryon, $1/2^-$ baryon, 
$S=-2$ isoscalar-scalar $H$ dibaryon~\cite{Yamaguchi:2016kxa}, 
scalar meson, pseudoscalar meson, and vector meson, respectively. 
The coupling constant $g_{BB'H}$ is given by 
$g_{\Lambda\Lambda H}=-g/\sqrt{8}$ and 
$g_{N\Xi H}=-g/\sqrt{2}$, where $g\approx2.4/\sqrt{\mathrm{MeV}}$ 
is used to reproduce the HAL-QCD collaboration results 
in the flavor SU(3) limit~\cite{Yamaguchi:2016kxa,Inoue:2011ai}. 
The values of $g_{(S,P,V)BB}$ are obtained 
from the Nijmegen soft-core model (NSC)~\cite{Stoks:1999bz}, 
whereas those for $g_{PBB'}$ are determined by the experimental 
results~\cite{pdg} and the SU(3) relations. 
All the relevant couplings for $\Lambda p\to K^+\Lambda\Lambda$ 
and $\Lambda p \to K^+\Xi^-p$
are listed in Table~\ref{tab:tab1}.
\begin{table}[!h]
{\renewcommand{\arraystretch}{1.2}
\begin{tabular}{c||c|c|c|c|c} \hline
$h$ 
& $\sigma$ & $\eta$ & $\omega$ & $\kappa$ & $K$ \\ \hline
$\Gamma_h$ [MeV] & $550$ & $1.31$ & $8.49$ & $478$ & $0$ \\ \hline\hline
$h$ & $K^*$\hspace{0.5cm} &\multicolumn{3}{c|}{$(N,\Lambda,\Xi)$} &\hspace{0.2cm}$N^*(1650)$\hspace{0.2cm}\\ \hline
$\Gamma_h$ [MeV]& $50.8$& \multicolumn{3}{c|}{$0$} &$158.2$ \\ \hline
\end{tabular}
}
\caption{Values of the full decay widths for the relevant hadrons from Refs.~\cite{pdg}.}
\label{tab:tab1}
\end{table}

Diagrams $(a)$ and $(b)$ for the $\Lambda p\to K^+\Lambda\Lambda$ 
($\Lambda\Lambda$ channel in the following) 
can be computed using these Lagrangians, resulting in the following invariant amplitudes:
\begin{eqnarray}
\label{eq:AMPAB}
i\mathcal{M}^{\Lambda\Lambda}_{(a)}\hskip-0.1cm&=&g_{KN\Lambda}g^2_{\Lambda\Lambda H}
u^\dagger_5[u^\dagger_4\,D^0_H(q_{4+5})u_1]D^{1/2}_\Lambda(q_{2-3})\gamma_5u_2 \nonumber \\
&-&(4\leftrightarrow5), \\
i\mathcal{M}^{\Lambda\Lambda}_{(b)}&=&\hskip-0.1cm\frac{1}{\sqrt{2}}g_{K\Lambda\Xi}g_{N\Xi H}g_{\Lambda\Lambda H}
u^\dagger_4u^\dagger_5\,D^0_H(q_{4+5})\,u_2\,D^{1/2}_\Xi(q_{1-3})\gamma_5u_1 \nonumber \\
&-&(4\leftrightarrow5),
\end{eqnarray}
where $q_{i\pm j}\equiv k_i\pm k_j$ and $D^s_h$ indicate
the \textit{dressed} propagator 
for a hadron $h$ with spin $s$. Its explicit form in the present work 
is as follows:
\begin{eqnarray}
\label{eq:PROP}
D^0_h(q)&=&\frac{F_h(q^2)}{q^2-M^2_h-i\Gamma_h M_h}, \\
D^1_{h\mu\nu}(q)&=&\frac{F_h(q^2)}{q^2-M^2_h-i\Gamma_h M_h}
\left(g_{\mu\nu}-\frac{q_\mu q_\nu}{M^2_h}\right),\\
D^{1/2}_h(q)&=&\frac{(\rlap{/}{q}+M_h)F_h(q^2)}{q^2-M^2_h-i\Gamma_h M_h}.
\end{eqnarray}
Here, the factors $M_h$, $\Gamma_h$, and $q$, denote the mass 
and full decay width of the hadron $h$, 
and the transferred momentum, respectively. The values of $\Gamma_h$ are 
listed in Table~\ref{tab:tab1}. 
The phenomenological form factor $F_h$ 
presents the spatial extension of the hadron $h$. In this study, we employ the following type of the form factors: 
\begin{equation}
\label{eq:FF}
F_h(q^2)=\frac{\Lambda^4_h}{\Lambda^4_h+(M^2_h-q^2)^2}.
\end{equation}

%
\begin{widetext}
\begin{table*}[!htb]
\centering
{\renewcommand{\arraystretch}{1.2}
\begin{tabular}{c|c|c|c|c|c|c} \hline
$\Lambda\Lambda H$&$N\Xi H$
&$(\kappa,K,K^*)N\Lambda$
&$(\kappa,K,K^*)\Lambda\Xi$
&$(\sigma,\eta,\omega)NN$
&$(\sigma,\eta,\omega)\Lambda\Lambda$
&$(\sigma,\eta,\omega)\Xi\Xi$\\
\hline
$\frac{-1.2}{\sqrt{2\,\mathrm{MeV}}}$
&$\frac{-2.4}{\sqrt{2\,\mathrm{MeV}}}$
&$(-8.31,-13.4,-4.26)$
&$(8.77,3.92,4.26)$
&$(-0.50,6.34,10.4)$
&$(-6.45,-6.86,4.96)$
&$(-12.6,-11.1,1.95)$\\
\hline
\hline
$KN^*\Lambda$
&$\eta NN^*$
&$(\kappa,K,K^*)N\Sigma$
&$(\kappa,K,K^*)\Lambda\Xi$
&$(\kappa,K,K^*)\Sigma\Xi$
&$\pi\Lambda\Sigma$
&$\pi NN$\\
\hline
$0.53$&$0.35$
&$(-5.32,-4.09,-2.46)$
&$(8.77,3.92,4.26)$
&$(-4.54,-16.7,-2.46)$
&$11.9$
&$13.0$\\ \hline
\end{tabular}
}
\caption{Relevant coupling constants for the present reaction process. 
These values are obtained from Refs.
~\cite{Inoue:2011ai,Yamaguchi:2016kxa,Stoks:1999bz,Garzon:2014ida,An:2011sb}. 
Here, $\sigma$ and $N^*$ denote $f_0(500,0^+)$ and $N^*(1650,1/2^-)$, 
which is the most important contribution in the vicinity of the production threshold.}
\label{tab:tab2}
\end{table*}
\end{widetext}
The cutoff mass $\Lambda_h$ is 
determined from other experimental data in the next section. 
Notably, the interchange of the $\Lambda$ baryons 
in the final state $(4\leftrightarrow5)$ in Eq.~(\ref{eq:AMPAB}) 
gives a negative sign, owing to the Fermi-Dirac statistics. 
All the relevant meson-baryon couplings are obtained 
from the Nijmegen soft-core potential model~\cite{Stoks:1999bz}, 
as listed in Table~\ref{tab:tab2}.

The invariant amplitudes for the diagram ($c$) can be evaluated as follows:
\begin{widetext}
\begin{eqnarray}
\label{eq:AMPC}
i\mathcal{M}^{\sigma p}_{(c)}&=&-g_{\sigma\Lambda\Lambda }g_{KN\Lambda}g_{\sigma NN}
[\bar{u}_5\gamma_5 D^{1/2}_p(q_{3+5}) u_2] D^0_{\sigma}(q_{1-4})[\bar{u}_4 u_1]-(4\leftrightarrow5),
\\
i\mathcal{M}^{\eta p}_{(c)}&=&-g_{\eta\Lambda\Lambda }g_{KN\Lambda}g_{\eta NN}
[\bar{u}_5\gamma_5 D^{1/2}_p(q_{3+5}) \gamma_5u_2]D^0_\eta(q_{1-4}) [\bar{u}_4\gamma_5u_1]-(4\leftrightarrow5),
\\
i\mathcal{M}^{\omega p}_{(c)}&=&-g_{\omega\Lambda\Lambda }g_{KN\Lambda}g_{\omega NN}
[\bar{u}_5\gamma_5 D^{1/2}_p(q_{3+5}) \gamma^\mu (q_{1-4}) u_2]D^1_{\omega\mu\nu}
[\bar{u}_4\gamma^\nu u_1]-(4\leftrightarrow5).
\\
i\mathcal{M}^{\eta p^*}_{(c)}&=&-g_{\eta\Lambda\Lambda }
g_{KN^*\Lambda}g_{\eta NN^*}
[\bar{u}_5 D^{1/2}_{p^*}(q_{3+5})u_2] D^0_\eta(q_{1-4}) [\bar{u}_4\gamma_5u_1]-(4\leftrightarrow5),
\end{eqnarray}
\end{widetext}
for the $(\sigma,\eta,\omega)$ meson exchange in the $t$-channel. 
The superscripts in $i\mathcal{M}^{h_1h_2}_{(c)}$ 
denote the intermediate hadrons as shown in Fig. \ref{fig:fig1}. 
Regarding the nucleon-resonance and $\Delta$-baryon contributions, 
we only consider the couplings to the $\eta$ meson 
to avoid theoretical uncertainties. 

The scalar meson $\sigma$ represents $f_0(500,0^+)$ ~\cite{pdg}. 
For the production of $H$-dibaryon near the threshold 
($\sqrt{s}\approx 2725$ MeV), 
only the nucleon resonance $N^*(1650,1/2^-)$ becomes relevant 
to the amplitude, $i\mathcal{M}^{\eta p^*}_{(c)}$. 
The strong coupling constants corresponding to $N^*(1650,1/2^-)$ 
are also obtained from the chiral coupled-channel method~\cite{Garzon:2014ida}, 
as listed in Table~\ref{tab:tab2}. 
Similarly, the $(\kappa,K,K^*)$ meson-exchange contributions are as follows:
\begin{widetext}
\begin{eqnarray}
\label{eq:AMPD}
\mathcal{M}^{\kappa^- p}_{(d)}&=&-g_{KN\Lambda}g^2_{\kappa N\Lambda} 
[\bar{u}_4\gamma_5 D^{1/2}_p(q_{3+4})u_1]D^0_{\kappa}(q_{5-2}) [\bar{u}_5 u_2]-(4\leftrightarrow5),
\\
i\mathcal{M}^{K^-p}_{(d)}&=&-g_{KN\Lambda }g^2_{KN\Lambda }
[\bar{u}_4\gamma_5 D'^{1/2}_p(q_{3+4}) \gamma_5 u_1]D^0_{K^-}(q_{5-2})
[\bar{u}_5\gamma_5u_2]-(4\leftrightarrow5),
\\
\mathcal{M}^{K^{*-} p}_{(d)}&=&-g_{KN\Lambda}g^2_{K^*N\Lambda}
[\bar{u}_4\gamma_5 D^{1/2}_p(q_{3+4}) \gamma^\mu u_1]
D^1_{K^*\mu\nu}(q_{5-2}) [\bar{u}_5\gamma^\nu u_2]-(4\leftrightarrow5)
\\
i\mathcal{M}^{K^- p^*}_{(d)}&=&-g_{KN\Lambda }g^2_{KN^*\Lambda}
[\bar{u}_4 D^{1/2}_{p^*}(q_{3+4})u_1]D^0_{K^-}(q_{5-2}) 
[\bar{u}_5\gamma_5u_2]-(4\leftrightarrow5),
\end{eqnarray}
\end{widetext}
where the strange scalar meson denotes $\kappa^-(800)$~\cite{pdg}.

The background contributions, which do not form resonant band structures 
in the Dalitz plot, are given by the following:
\begin{widetext}
\begin{eqnarray}
\label{eq:AMPEF}
i\mathcal{M}^{\sigma\Lambda}_{(e)}&=&-g_{KN\Lambda}g^2_{\sigma\Lambda\Lambda}
[\bar{u}_5D^{1/2}_\Lambda(q_{2-3})\gamma_5u_2]
D^0_{\sigma}(q_{1-4})[\bar{u}_4u_1]-(4\leftrightarrow5),
\\
i\mathcal{M}^{\eta\Lambda}_{(e)}&=&-g_{KN\Lambda}g^2_{\eta\Lambda\Lambda}
[\bar{u}_5\gamma_5D^{1/2}_\Lambda(q_{2-3})\gamma_5u_2]
D^0_{\eta}(q_{1-4})[\bar{u}_4\gamma_5u_1]
-(4\leftrightarrow5),
\\
i\mathcal{M}^{\omega\Lambda}_{(e)}&=&-g_{KN\Lambda}g^2_{\omega\Lambda\Lambda}
[\bar{u}_5\gamma_5\gamma^\mu D^{1/2}_\Lambda(q_{2-3})\gamma_5u_2]
D^1_{\omega,\mu\nu}(q_{1-4})[\bar{u}_4\gamma^\nu u_1]
-(4\leftrightarrow5),
\end{eqnarray}
and
\begin{eqnarray}
\label{eq:AMPEF}
i\mathcal{M}^{\kappa^-p}_{(f)}&=&-g_{\kappa N\Lambda}g_{K\Lambda\Xi}g_{\kappa\Lambda\Xi}
[\bar{u}_4D^{1/2}_{\Xi^-}(q_{1-3})\gamma_5u_1]D^0_{\kappa^-}(q_{5-2})
[\bar{u}_5u_2]-(4\leftrightarrow5),
\\
i\mathcal{M}^{K^-p}_{(f)}&=&-g_{K\Lambda N}g_{K\Lambda\Xi}g_{K\Lambda\Xi}
[\bar{u}_4\gamma_5D^{1/2}_{\Xi^-}(q_{1-3})\gamma_5u_1]
D^0_{K^-}(q_{5-2})
[\bar{u}_5\gamma_5u_2]-(4\leftrightarrow5),
\\
i\mathcal{M}^{K^{*-}p}_{(f)}&=&-g_{K^*N\Lambda}g_{K\Lambda\Xi}g_{K^*\Lambda\Xi}
[\bar{u}_4\gamma^\mu D^{1/2}_{\Xi^-}(q_{1-3})\gamma_5u_1]
D^1_{K^*\mu\nu}(q_{5-2})[\bar{u}_5\gamma^\nu u_2]-(4\leftrightarrow5).
\end{eqnarray}
\end{widetext}
All the relevant coupling constants are provided in Table~\ref{tab:tab2}. 

For the $\Lambda p \to K^+\Xi^-p$ reaction, 
we can compute the invariant amplitudes similarly 
without the particle-exchange $(4\leftrightarrow5)$ terms in the final state. 
The relevant coupling constants for this reaction are provided in Table~\ref{tab:tab2}. 
In this reaction, the $27$-plet $\Theta^{++}$ pentaquark contribution can be 
considered in diagram (c). 
However, the existence of this exotic baryon has never 
been confirmed experimentally. 
Hence, we ignore this contribution for brevity. 
In diagrams ($d$) and ($e$), there are two baryon-pole contributions, 
that is, $\Lambda$ and $\Sigma^0$, which differ from the $\Lambda\Lambda$ channel. 

Unlike the electromagnetic hadron production involving 
the Ward-Takahashi identity, 
determining the phase factors between strong-interaction amplitudes
is a relatively difficult task owing to the lack of symmetry. In the present calculation, we employ a free parameter for the phase difference 
between the tree-level invariant amplitudes as follows: 
\begin{equation}
\label{eq:TAMP}
i\mathcal{M}_\mathrm{tree}=i\mathcal{M}_{(a+b)}
+e^{i\phi}\left[i\mathcal{M}_{(c+d+e+f)}\right].
\end{equation}
The phase factors among the invariant amplitudes of diagrams ($c$--$f$), 
except for the nucleon-resonance contributions, 
are determined from the coupling constants in the 
Nijmegen model~\cite{Stoks:1999bz}. 
Although we do not have a theoretical reasoning for fixing the phase factor 
for the nucleon-resonance contributions, we simply assume that 
$g_{MBN^*}$ is positively real. 
Moreover, the nucleon-resonance contributions 
were numerically verified to be negligible, 
owing to its significant full decay width as provided in Table~\ref{tab:tab1}. 
Thus, we introduce a single phase factor between the $H$-dibaryon contributions 
and others, as shown in Eq.~(\ref{eq:TAMP}). 
The phase angle $\phi$ is considered a free parameter 
in the numerical calculations. 

After attaining the aforementioned, the final-state interaction 
(FSI) contributions can be considered. 
The total amplitude including the FSI contributions is defined 
in the on-shell approximation (OnF)~\cite{Nam:2003ch} as follows:
\begin{eqnarray}
\label{eq:CC}
i\mathcal{M}^{\Lambda\Lambda,\mathrm{OnF}}_\mathrm{tree+FSI}&=&i\mathcal{M}^{\Lambda\Lambda}_\mathrm{tree}\\ \nonumber
&+&\underbrace{( i\mathcal{M}^{\Lambda\Lambda}_\mathrm{tree})\left(i\int\frac{d^4q}{(2\pi)^4}
\mathcal{G}^\mathrm{OnF}_{\Lambda\Lambda}\right)
(i\hat{\mathcal{M}}^\mathrm{cc,OnF}_{\Lambda\Lambda\to\Lambda\Lambda})}_\mathrm{FSI},
 \end{eqnarray}
where $i\hat{\mathcal{M}}^\mathrm{cc,OnF}_{B_1B_2\to B_3B_4}$ 
stands for the flavor-singlet two-baryon coupling constant 
for the $I=0$ and $S=-2$ channels in the coupled-channel (cc) method. 
Based on the isospin symmetry, its elementary amplitude can be expressed
as follows \cite{Sasaki:2015ifa}:
\begin{equation}
\label{eq:EL}
i\hat{\mathcal{M}}_{B_1B_2\to B_3B_4}=-i\lambda_1\left(
\begin{array}{ccc}
+\frac{1}{8}&-\frac{1}{2}&-\frac{\sqrt{3}}{4}\\
-\frac{1}{2}&+\frac{1}{2}&+\frac{\sqrt{3}}{2}\\
-\frac{\sqrt{3}}{4}&+\frac{\sqrt{3}}{2}&+\frac{1}{8}
\end{array}
\right).
\end{equation}
The value of $\lambda_1$ to reproduce the binding energy of 
the HAL QCD data  is given by $-12.8/\mathrm{GeV}^2$ ~\cite{Sasaki:2015ifa}. 
Note, in Eq.~(\ref{eq:CC}), only the baryon-baryon re-scattering
for FSI is considered for simplicity, 
and the $S=0$ meson-baryon re-scattering is ignored 
because we are interested in the baryon-baryon invariant mass spectrum. 
The two-baryon propagator $\mathcal{G}^\mathrm{OnF}_{B_1B_2}$ applies
the on-shell factorization. 
The integration of $\mathcal{G}^\mathrm{OnF}_{B_1B_2}$ 
over the loop momentum $q$ can be regularized simply 
using the dimensional-regularization method~\cite{Nam:2003ch} as follows: 
\begin{eqnarray}
\label{eq:}
G^\mathrm{PV,OnF}_{B_1B_2}
&\approx&
i\int\frac{d^4q}{(2\pi)^4}\frac{M_a(M_{4+5}-M_a+M_b)}
{[q^2-M^2_a-i\epsilon][(q_{4+5}-q)^2-M^2_b-i\epsilon]} \nonumber \\
&=& \frac{M_a(M_{4+5}-M_a+M_b)}{16\pi^2} \nonumber \\
&\times&\Bigg[\ln\frac{M^2_b}{\mu^2}+\frac{M^2_a-M^2_b+M^2_{4+5}}{2M^2_{4+5}}
\ln\frac{M^2_a}{M^2_b} \nonumber \\
&{}&~~ +\frac{\eta}{M_{4+5}}\left(L_{+-}+L_{++}-L_{-+}-L_{--}\right)\Bigg]
\end{eqnarray}
where we define
\begin{eqnarray}
\label{eq:}
\eta &\equiv& \frac{\sqrt{[M^2_{4+5}-(M_a-M_b)^2][M^2_{4+5}-(M_a+M_b)^2]}}{2M_{4+5}},\nonumber \\
L_{\pm\pm} &\equiv& \ln[\pm M^2_{45}\pm(M^2_b-M^2_a)+2M_{4+5}\eta].
\end{eqnarray}

Hence, in terms of the on-shell factorization, 
the coupled-channel amplitude for the $B_1B_2\to B_3B_4$ channel reads:
\begin{eqnarray}
\label{eq:VGT}
i\hat{\mathcal{M}}^\mathrm{cc,OnF}_{B_1B_2\to B_3B_4} &=& i\hat{\mathcal{M}}_{B_1B_2\to B_3B_4} \nonumber \\
&+&\sum_{B_\ell B_{\ell'}}(i\hat{\mathcal{M}}_{B_1B_2\to B_\ell B_{\ell'}}) \nonumber \\
&\times&(G^\mathrm{OnF}_{B_\ell B_{\ell'}})(i\hat{\mathcal{M}}_{B_\ell B_{\ell'}\to B_3B_4})+\cdots.
\end{eqnarray}
Owing to the regularization and Lorentz structure of the elementary amplitudes 
as shown in Eq.~(\ref{eq:EL}), 
each term in the r.h.s. of Eq.~(\ref{eq:VGT}) is finite. 
Eq.~(\ref{eq:VGT}) can therefore be rewritten in a matrix form as follows:
\begin{equation}
\label{eq:}
i\hat{\mathrm{M}}^\mathrm{cc,OnF}=
\left[\mathrm{I}_{3\times3}-(i\hat{\mathrm{M}})(\mathrm{G}^\mathrm{OnF})\right]^{-1}(i\hat{\mathrm{M}}).
\end{equation}
Here, channels (1,2,3) are defined by 
the $(\Lambda\Lambda,N\Xi,\Sigma\Sigma)$ scattering states. 

\begin{figure}[t]
\includegraphics[width=8cm]{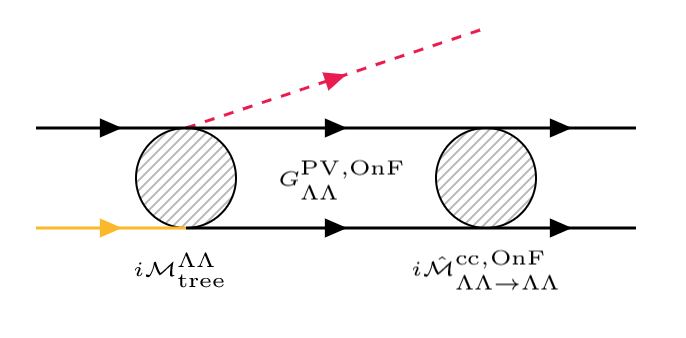}
\caption{(Color online) Schematic 
for the final-state interaction contribution in Eq.~(\ref{eq:FSILL}), 
for the $\Lambda\Lambda$ channel.}        
\label{fig:fig3}
\end{figure}

Regarding the $\Lambda\Lambda\to\Lambda\Lambda$ scattering for FSI, 
we consider only $i\mathrm{M}^\mathrm{cc}_{11}
=i\mathcal{M}^\mathrm{cc}_{\Lambda\Lambda\to\Lambda\Lambda}$ in Fig. \ref{fig:fig3}.
Following extensive calculations, the analytical form is obtained as   
follows:
\begin{widetext}
\begin{equation}
\label{eq:FSILL}
i\hat{\mathcal{M}}^\mathrm{cc,OnF}_{\Lambda\Lambda\to\Lambda\Lambda}
=-\frac{\lambda_1[16-6i\lambda_1(4G_2+3G_3)-15\lambda^2_1G_2G_3]}
{128i+16\lambda_1(G_1+4G_2+3G_3)+6i\lambda^2_1(4G_1G_2+3G_1G_3
+12G_2G_3)-15\lambda^3_1G_1G_2G_3},
\end{equation}
\end{widetext}
where $G_{1,2,3}$ indicates $G_{\Lambda\Lambda,N\Xi,\Sigma\Sigma}$. 
Finally, considering all the factors previously indicated, 
we obtain the FSI-corrected total amplitude for the $\Lambda\Lambda$ channel:
\begin{eqnarray}
\label{eq:CCON}
i\mathcal{M}^{\Lambda\Lambda,\mathrm{OnF}}_\mathrm{tree+FSI}
&=&i\mathcal{M}^{\Lambda\Lambda}_\mathrm{tree}\left[1+
(G^\mathrm{PV,OnF}_{\Lambda\Lambda})(i\hat{\mathcal{M}}^\mathrm{cc,OnF}_{\Lambda\Lambda\to\Lambda\Lambda})\right].
 \end{eqnarray}
Similarly, we can derive the $\Xi^-p$ channel total amplitude as follows:
\begin{eqnarray}
\label{eq:CCON2}
i\mathcal{M}^{\Xi^-p,\mathrm{OnF}}_\mathrm{tree+FSI}
&=&i\mathcal{M}^{\Xi^-p}_\mathrm{tree}
\left[1+\frac{1}{\sqrt{2}}(G^\mathrm{PV,OnF}_{\Xi^-p})(i\hat{\mathcal{M}}^\mathrm{cc,OnF}_{\Xi^-p\to\Xi^-p})\right],
 \end{eqnarray}
where
\begin{widetext}
\begin{equation}
\label{eq:}
i\hat{\mathcal{M}}^\mathrm{cc,OnF}_{\Xi^-p\to\Xi^-p}
=-\frac{\lambda_1[64+24i\lambda_1(4G_2+3G_3)-15\lambda^2_1G_2G_3]}
{128i+16\lambda_1(G_1+4G_2+3G_3)+6i\lambda^2_1(4G_1G_2+3G_1G_3
+12G_2G_3)-15\lambda^3_1G_1G_2G_3}.
\end{equation}
\end{widetext}
\section{Numerical results and Discussions}
In this section, we provide the numerical calculation results 
with details regarding the $\Lambda p\to K^+\Lambda\Lambda$ 
($\Lambda\Lambda$ channel) and $\Lambda p\to K^+\Xi^-p$ ($\Xi^-p$ channel) 
reaction processes. In this calculation, the $H$-dibaryon is assumed to be
unbound above the $\Lambda\Lambda$ threshold.   
The mass range of the $H$-dibaryon is strongly connected with 
the observation of the double $\Lambda$ hypernuclei, 
which imposes that the $H$-dibaryon
mass should be larger than 2.22 GeV$/c^2$. 
Recent Lattice QCD calculation results indicate that
the mass ranges between $\Lambda\Lambda$ and $\Xi^-p$ thresholds
\cite{sasaki,Inoue:2011ai,Haidenbauer:2011ah,Shanahan:2011su,Yamaguchi:2016kxa}. 
Two $H$-dibaryon states, below and above the $\Xi^-p$ threshold, 
were chosen
considering the $H(2250)\to\Lambda\Lambda$ and $H(2270)\to \Xi^-p$ decays.

As indicated in Section II, 
we employ the coupling constants for the dibaryon $g_{BB'H}$ 
from the bare $H$-dibaryon model, in which the values were determined to fit 
the flavor SU(3) symmetric HAL-QCD data~\cite{Yamaguchi:2016kxa}. 
Therefore, the values of $g_{BB'H}$ may be different from reality, 
where the flavor symmetry is heavily broken. Nonetheless, 
as guidance for the present theoretical calculations, 
these symmetric values were adopted as a trial. 
In Ref.~\cite{Inoue:2011ai}, 
the full decay width of the dibaryon was 
$\Gamma_H=2.7$ MeV at the physical point. 

\begin{figure}[hbt]
\includegraphics[width=7cm]{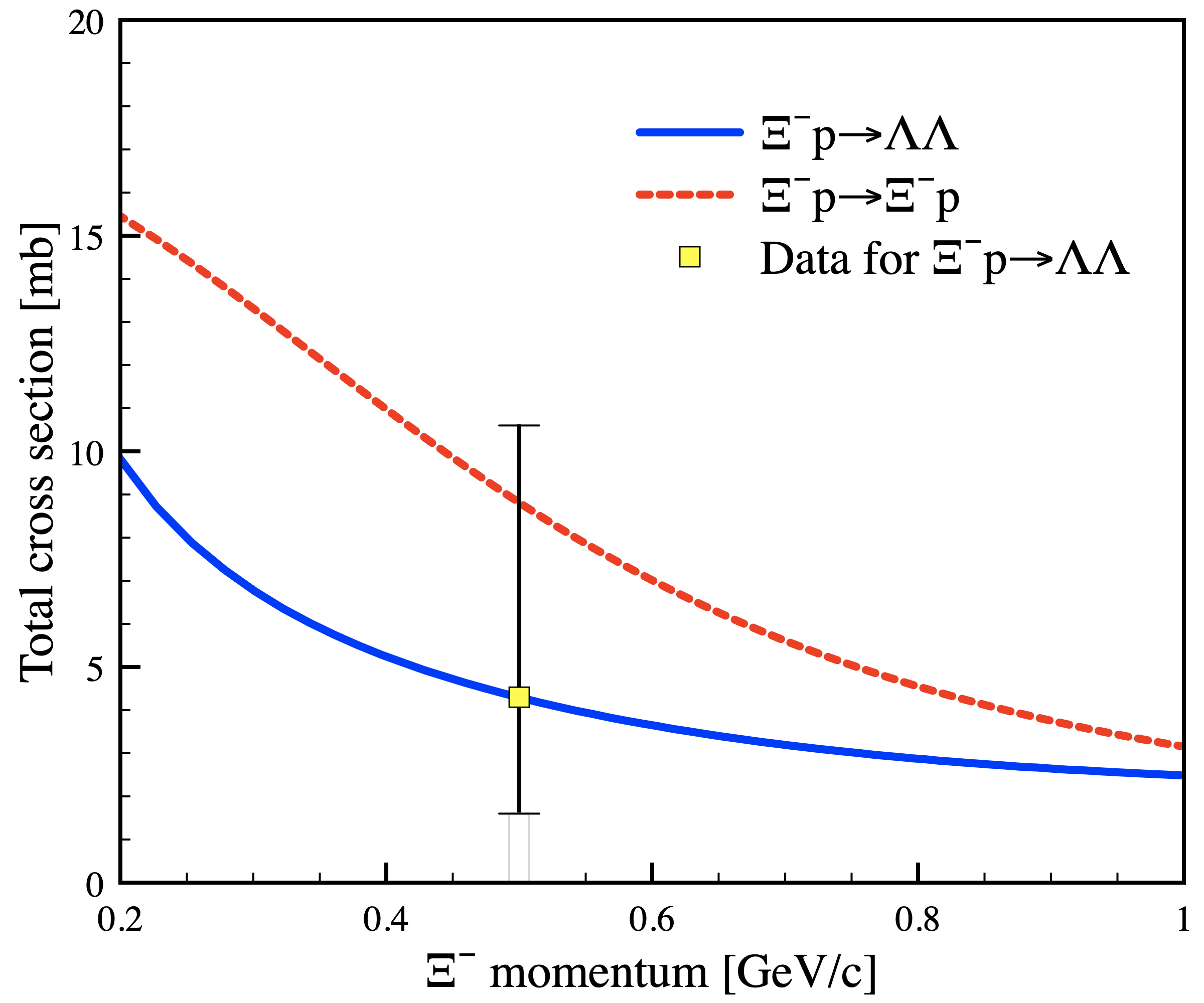}
\caption{(Color online) Total cross-section for $\Xi^-p\to\Lambda\Lambda$ (solid)
and $\Xi^-p\to\Xi^-p$ 
(dotted) using Eqs.~(\ref{eq:XPLL1}) and (\ref{eq:XPLL2}) with the cutoff mass 
$\Lambda_{\Lambda\Lambda}=435$ MeV 
and $\Lambda_{\Xi^-p}=535$ MeV, respectively. $\Lambda_{\Lambda\Lambda}$ 
is determined to reproduce the data point $\sigma=4.3^{+6.3}_{-2.7}$, extracted 
from $^{12}$C$(\Xi^-,\Lambda\Lambda)X$ data~\cite{e224b}.}       
\label{fig:fig4}
\end{figure}

First, the cutoff mass was fixed in the form factors in Eq.~(\ref{eq:FF}). 
In Ref.~\cite{e224b}, a few events of the $^{12}$C$(\Xi^-,\Lambda\Lambda)X$ reaction were reported.
Using the eikonal approximation, the total cross-section was deduced to 
$\sigma=4.3^{+6.3}_{-2.7}$ for the $\Xi^-p\to\Lambda\Lambda$ reaction at 
$p_{\Xi^-}=0.5$ GeV$/c$. 
We reproduced this value in the present theoretical framework.
For simplicity, we only considered the $\kappa$, $K$, and $K^*$ exchanges 
in the $t$-channel, and ignored a possible $H$-dibaryon contribution 
in the $s$-channel. Moreover, the cutoff masses for the three meson exchanges were chosen to be the same for brevity.
The relevant invariant amplitude is then obtained as follows:
\begin{equation}
\label{eq:XPLL1}
i\mathcal{M}_{\Xi^-p\to\Lambda\Lambda}=\sum_{\Phi=\kappa,K,K^*}i\mathcal{M}_{\Phi}-(c\leftrightarrow d),
\end{equation}
where
\begin{eqnarray}
\label{eq:XPLL2}
i\mathcal{M}_{\Phi} &=& \frac{ig_{\Phi\Xi\Lambda}g_{\Phi N\Lambda}(\bar{u}_c\Gamma_\Phi  u_a)
(\bar{u}_d\Gamma^\Phi u_b)F_\Phi(t)}{t-M^2_\Phi-i\Gamma_\Phi M_\Psi},\nonumber \\
\Gamma_{\kappa,K,K^*} &=& (\bm{1}_{4\times4},\gamma_5,\gamma_\mu).
\end{eqnarray}
All relevant inputs are listed in Tables~\ref{tab:tab1} and \ref{tab:tab2}. 
Thus, 
the cutoff mass is determined to be $\Lambda=435$ GeV. 
The numerical results are shown in Fig.~\ref{fig:fig4}.

\begin{figure}[t]
\centering
\stackinset{r}{1.5cm}{t}{0.5cm}{(a)}{
\includegraphics[width=8cm]{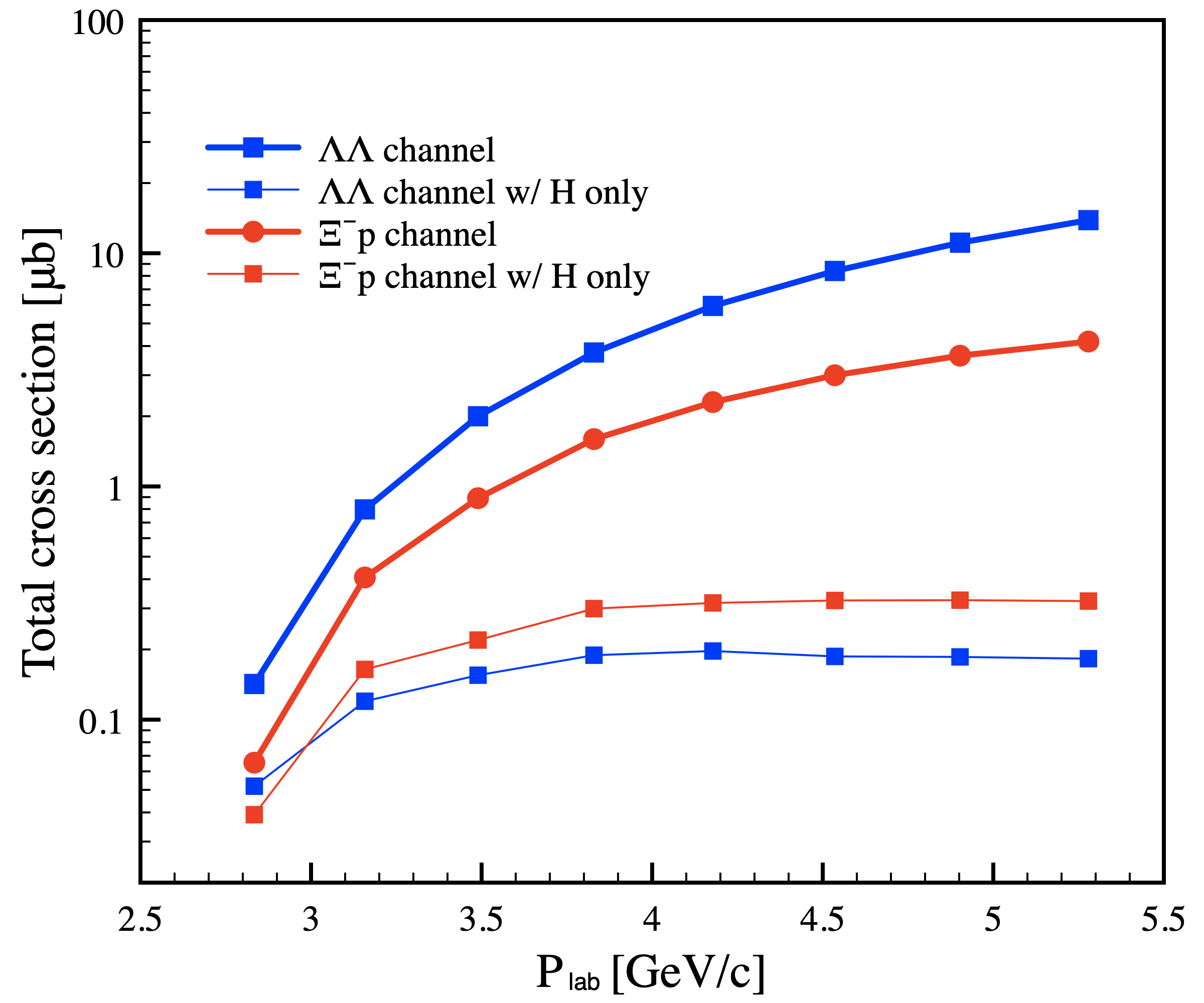}
}
\stackinset{l}{1.5cm}{t}{0.5cm}{(b)}{
\includegraphics[width=8cm]{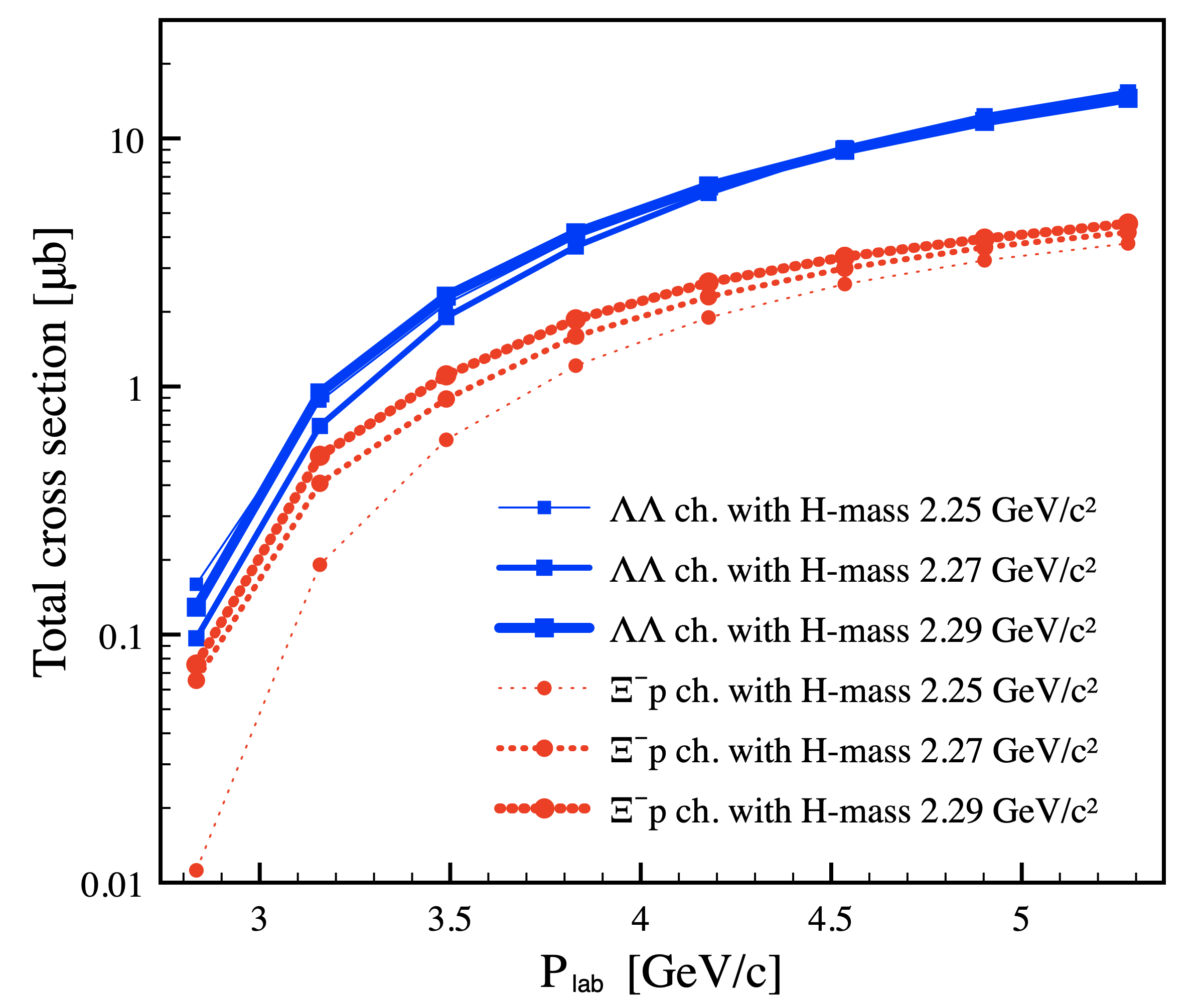}
}
\caption{(Color online) (a) Total cross-sections for the $\Lambda\Lambda$ (square) and  $\Xi^-p$ (circle) channels for the total (thick) and $H$-dibaryon-only (thin) contributions as functions of the $\Lambda$ beam momentum $p_\mathrm{lab}$. 
(b) Total cross-sections by varying the masses of the $H$ dibaryon.}  
\label{fig:fig5}
\end{figure}

In the left panel of Fig.~\ref{fig:fig5}, the numerical results 
for the total cross-sections of the $\Lambda\Lambda$ (square) 
and  $\Xi^-p$ (circle) channels are presented for the total (thick) 
and $H$-dibaryon-only (thin) contributions as a function of 
the $\Lambda$ beam momentum $p_\mathrm{lab}$. 
We determined that the total cross-sections for the two channels are 
of the order of approximately a few $\mu$b, 
which is smaller than that for 
the $pp\to K^+\Lambda p$ from the COSY experiment~\cite{Balewski:1991ns}. 
The total cross-sections from the $\Lambda\Lambda$ channel 
are approximately twice as large as that of the $\Xi^-p$
because there are more possible contributions, 
as shown in the relevant Feynman diagrams in Fig.~\ref{fig:fig1}, 
in addition to the larger Nijmegen coupling constants. On the contrary, if we only consider the $H$-dibaryon, 
the order of the cross-sections is reversed, 
owing to the value of $g_{H\Lambda\Lambda}$ 
being smaller than $g_{H\Xi^-p}$ by a factor of two 
considering the isospin factor. Note, the production cross-section for the $H$ dibaryon 
is \textit{a few tens of nanobarn}. 
As shown in the right panel of Fig~\ref{fig:fig5},
to test the $H$-dibaryon mass dependence of the total cross-sections, 
they are depicted with 
$M_H=(2.25\sim2.29)\,\mathrm{GeV}/c^2$ for the two reaction channels. 
The effects from the mass changes are unapparent, 
while considerable difference can be observed 
for the $\Xi^-p$ channel with $M_H=2.25\,\mathrm{GeV}/c^2$.

\begin{figure}[t]
\centering
\stackinset{l}{1.5cm}{t}{0.5cm}{(a)}{
\includegraphics[width=8cm]{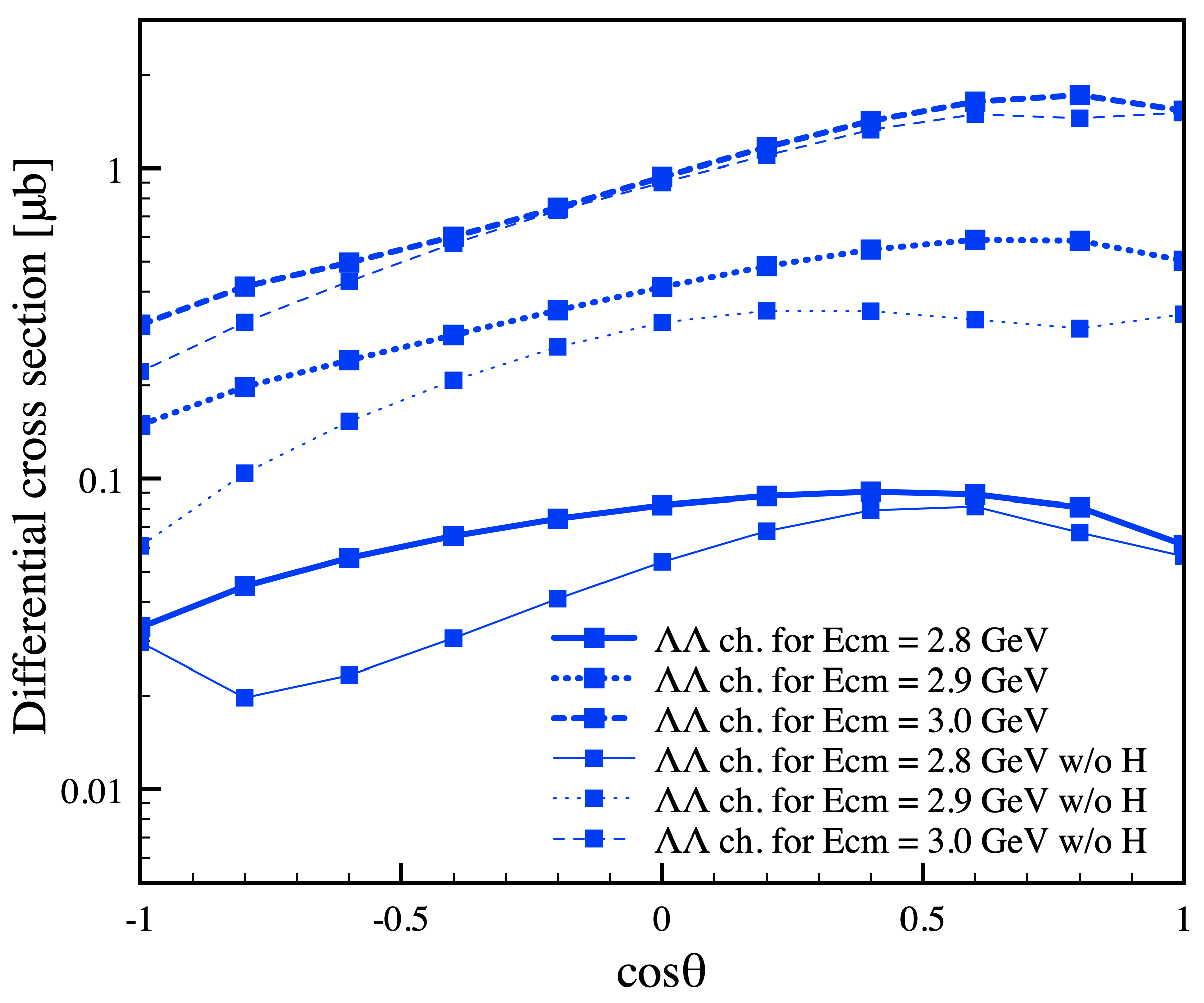}
}
\stackinset{l}{1.5cm}{t}{0.5cm}{(b)}{
\includegraphics[width=8cm]{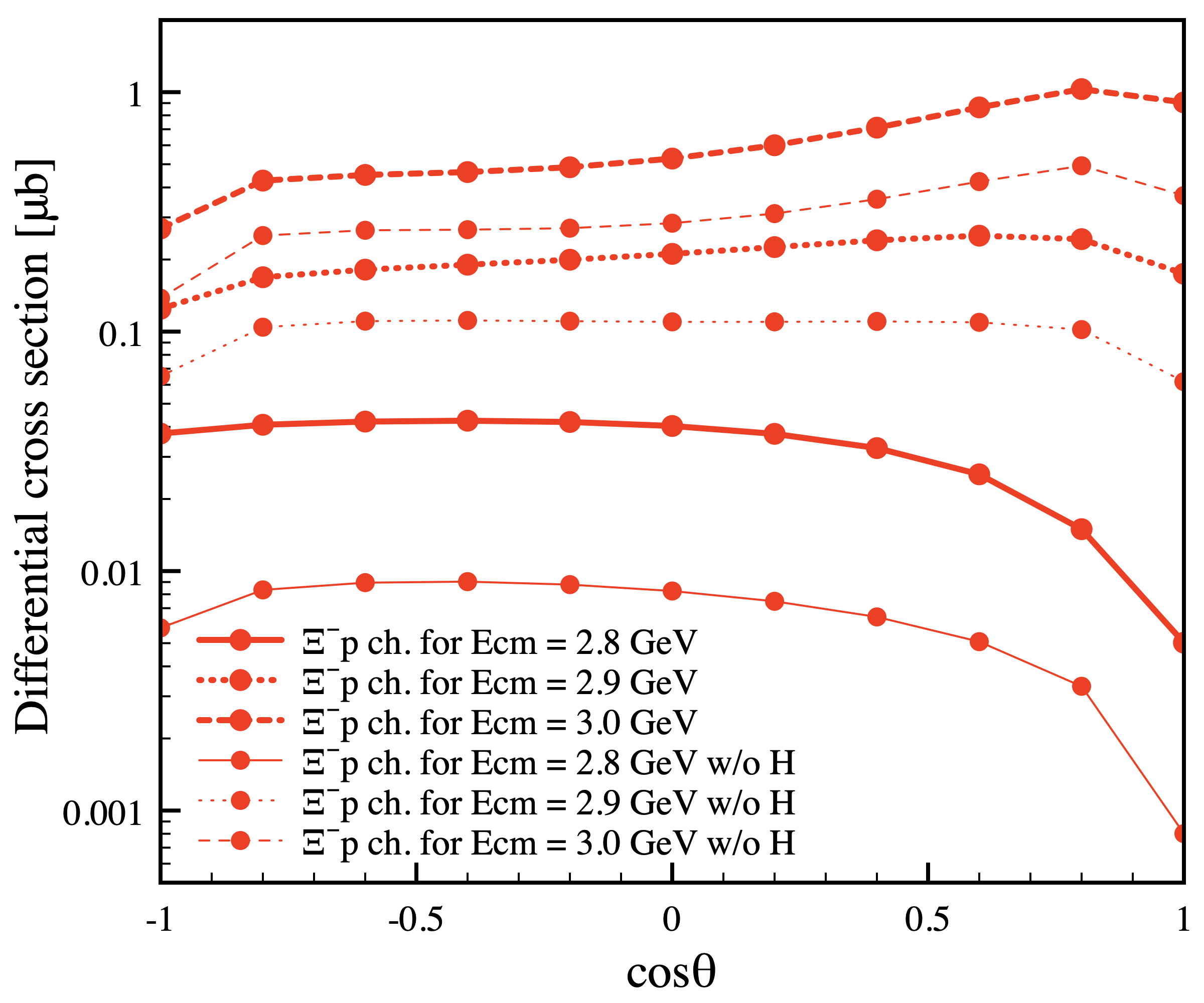}
}
\caption{(Color online) Differential cross-sections for (a) $\Lambda\Lambda$ 
and (b) $\Xi^-p$ (right) channels, respectively, as functions of 
the scattering angle of $K^+$ in the center-of-mass frame (cm) $\theta$.}  
\label{fig:fig6}
\end{figure}

In Fig~\ref{fig:fig6}, 
the numerical results for the differential cross-sections 
of the $\Lambda\Lambda$ (left) and $\Xi^-p$ (right) channels
are presented as the function of the scattering angle of $K^+$ 
in the center-of-mass frame (cm) $\theta$. 
We also analyzed the differential cross-sections 
in the energy range of $E_\mathrm{cm}=2.8$--$3.0$ GeV. 
The thick and thin lines denote the cases 
with and without the $H$-dibaryon contributions, respectively. 
The angular dependence for the two channels is relatively flat 
at a low energy and forwarding as the energy increases. 
Note, the angular dependence of the $H$-dibaryon production 
is nearly flat at high energies, indicating the $S$-wave nature of the particle. 

\begin{figure}[t]
\centering
\stackinset{l}{-.1cm}{t}{0.2cm}{(a)}{
\includegraphics[width=8cm]{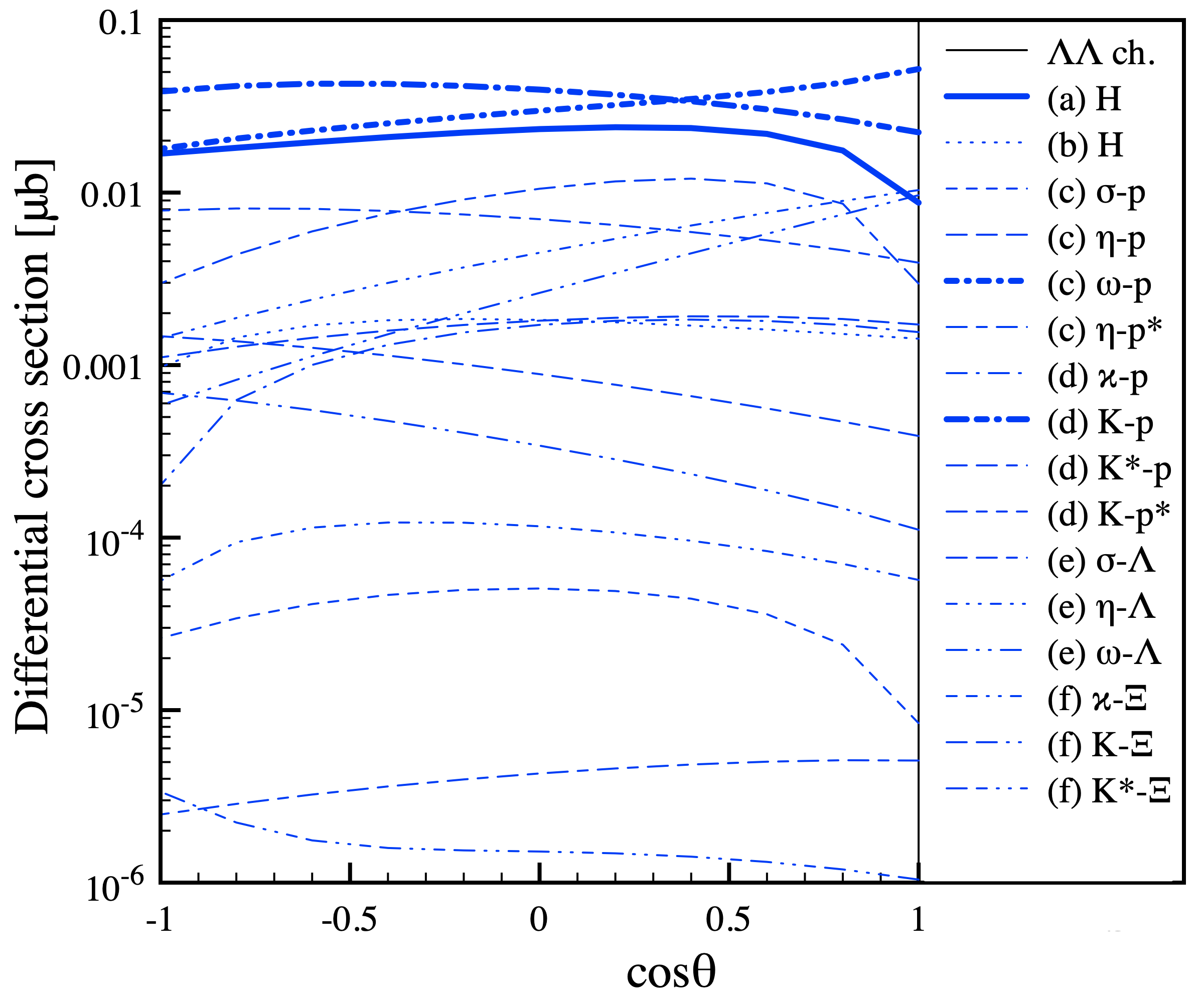}
}
\stackinset{l}{-.1cm}{t}{0.2cm}{(b)}{
\includegraphics[width=8cm]{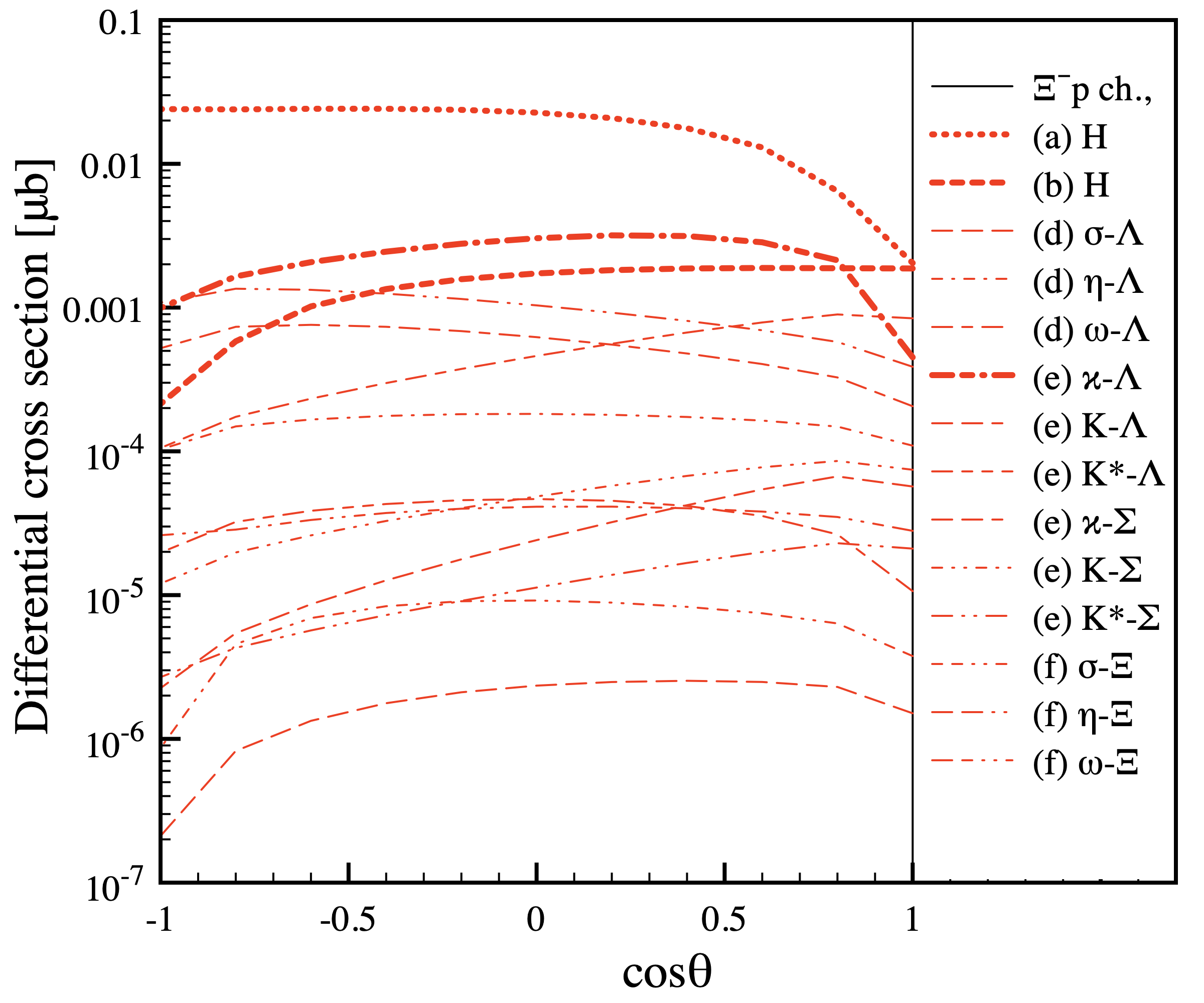}
}
\caption{(Color online) Differential cross-sections for 
(a) the $\Lambda p\to K^+\Lambda\Lambda$ and 
(b) $\Lambda p\to K^+\Xi^-p$ reactions 
at $E_\mathrm{cm}=2.8$ GeV ($p_\Lambda=2.83$ GeV$/c$) 
are plotted in separate curves for individual diagram contributions.}       
\label{fig:fig7}
\end{figure}

To investigate the production mechanisms more carefully, 
we present the numerical results for the differential cross-sections 
for each contribution individually
at $E_\mathrm{cm}=2.8$ GeV ($p_\mathrm{lab}=2.83$ GeV$/c$), 
in the same manner as presented in Fig~\ref{fig:fig7}. 
Regarding the $\Lambda\Lambda$ channel,  
the $\omega$ and $K^-$ exchanges in the proton-pole diagrams ($c$ and $d$)
are predominant owing to the combinations of 
the larger Nijmegen coupling constants. 
Moreover, the $H$-dibaryon production diagram 
with the $\Lambda$ pole ($a$) is significantly larger 
than that of the $\Xi^-$-pole diagram ($b$). 
Regarding the $\Xi^-p$ channel, 
the $H$-dibaryon production diagrams are considerably larger than others, 
and the $\kappa$ exchange in the $t$ channel 
($e$) provides a meaningful contribution. 
Generally, we determined that the $H$-dibaryon production diagram ($a$) 
with the Mandelstam variable $t=(k_\Lambda-k_{K^+})^2$ enhances 
forward scattering, and vice versa for the diagram ($b$) 
with $u=(k_p-k_{K^+})^2$, as expected. 
 
\begin{figure}[t]
\centering
\stackinset{r}{2.5cm}{t}{2.8cm}{(a)}{
\includegraphics[width=9.2cm]{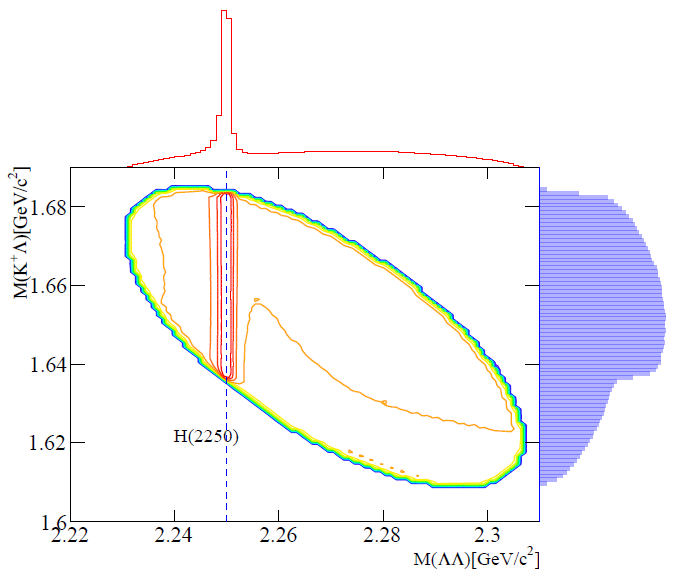}
}
\stackinset{r}{2.2cm}{t}{2.8cm}{(b)}{
\includegraphics[width=8.8cm]{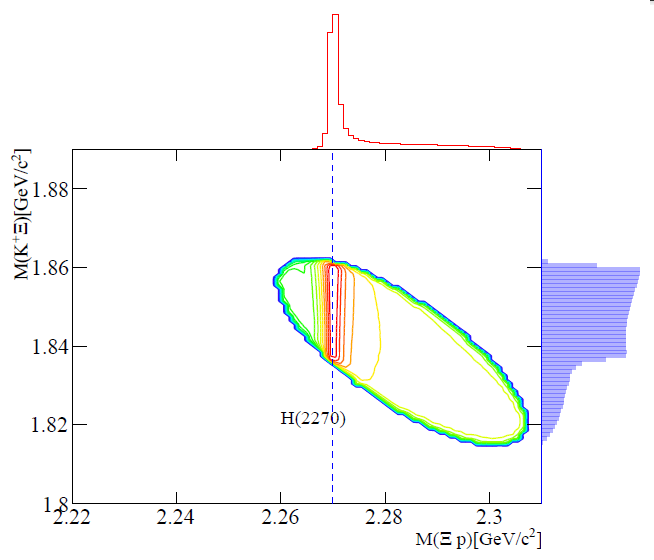}
}
\caption{(Color online) 
Dalitz plots for (a) the $\Lambda p\to K^+\Lambda\Lambda$ and (b) 
$\Lambda p\to K^+\Xi^-p$ reactions at $p_\Lambda=2.83$ GeV$/c$, respectively.
Both contain the relative phase angle $\psi=0$.
The Dalitz plots are projected onto the $\Lambda\Lambda/\Xi^-p$ and $K^+\Lambda$/
$K^+\Xi^-$ mass axes and plotted as histograms on the top and right sides, respectively. }     
\label{fig:fig8}
\end{figure}

\begin{figure}[t]
\centering
\stackinset{r}{.5cm}{t}{0.5cm}{(a)}{
\includegraphics[width=8cm]{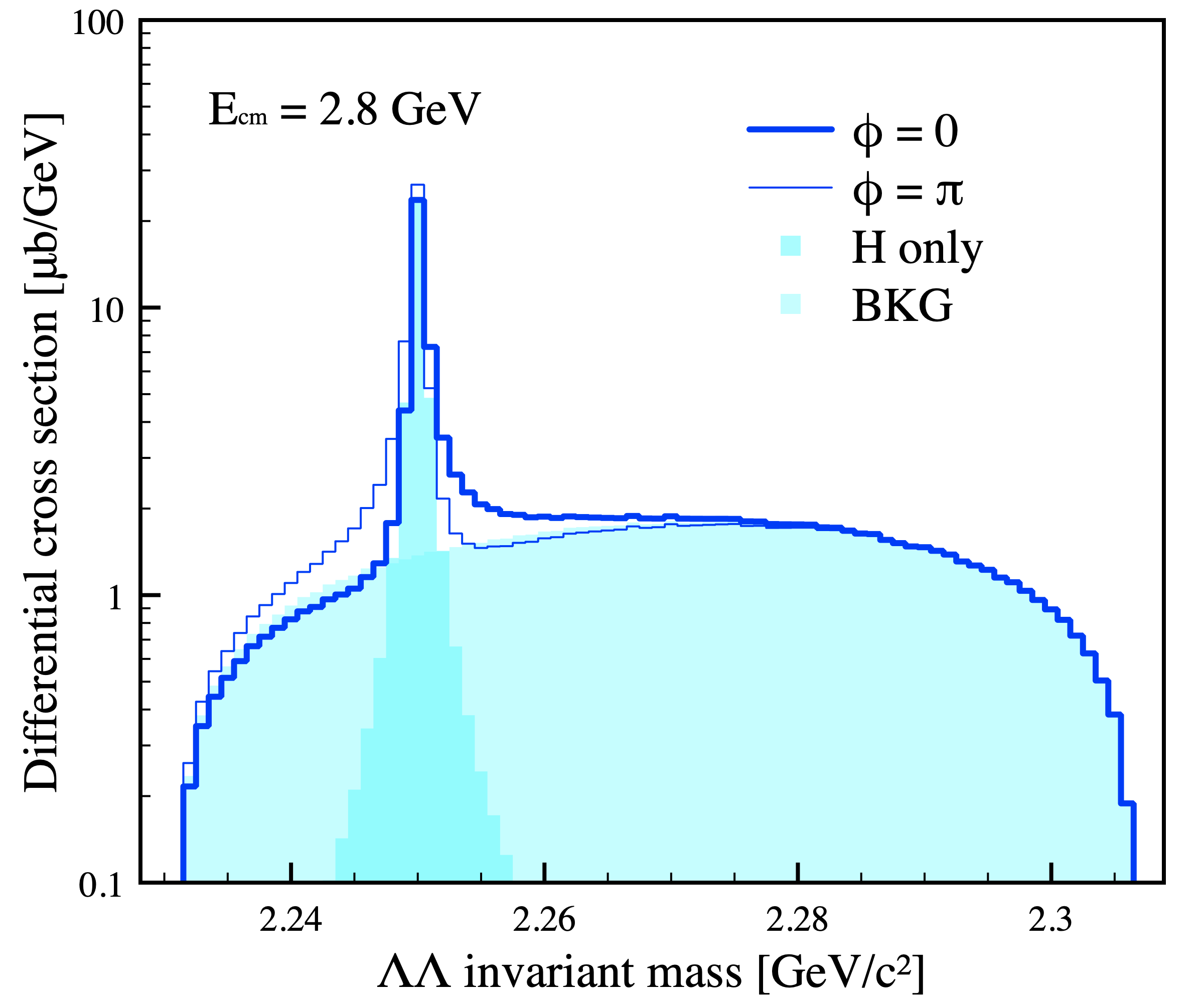}
}
\stackinset{r}{.5cm}{t}{0.5cm}{(b)}{
\includegraphics[width=8cm]{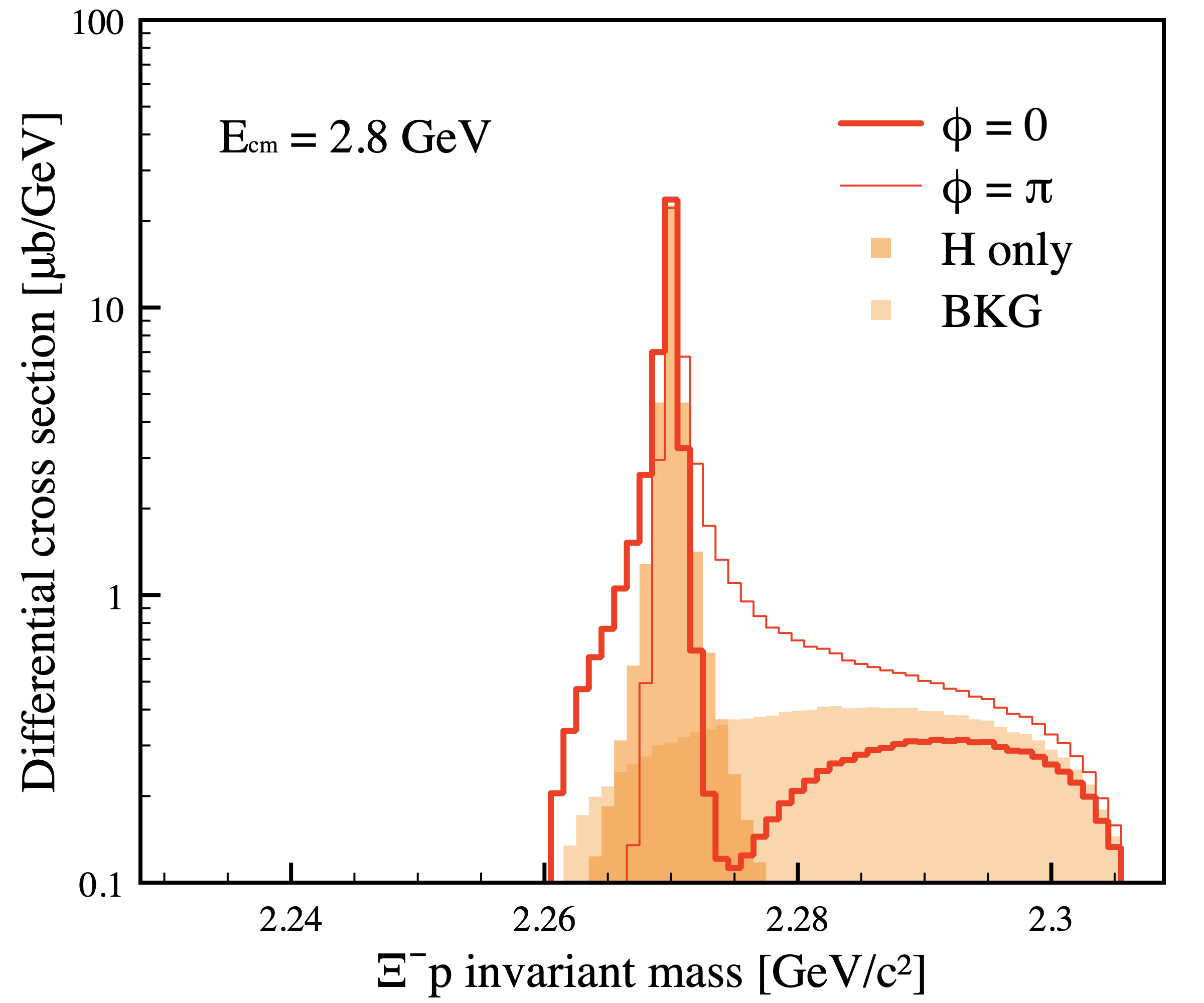}
}
\caption{(Color online) 
Invariant-mass plots for 
$M_H=2.25\,\mathrm{GeV}/c^2$ and $2.27\,\mathrm{GeV}/c^2$ for 
(a) $\Lambda\Lambda$ and (b) $\Xi^-p$ channels, respectively, 
at $E_\mathrm{cm}=2.8$ GeV. 
The width of the $H$ dibaryon is chosen to be $1$ MeV and 
the phase angle $\phi$ is tested for $0$ (thick) and $\pi$ (thin). 
The shaded areas indicate the cases without the $H$-dibaryon.
}     
\label{fig:fig9}
\end{figure}

The Dalitz plot for the $\Lambda p\to K^+\Lambda\Lambda$ and 
$\Lambda p\to K^+\Xi^-p$ reactions are plotted in Fig. \ref{fig:fig8}
for a $\Lambda$ beam momentum of 2.83 GeV$/c$. Because no background 
processes form structure in the Dalitz plots, the $H$-dibaryon band appears
predominant.
The numerical results for the invariant-mass plots are provided 
in Fig.~\ref{fig:fig9} with $M_H=2.25\,\mathrm{GeV}/c^2$ and 
$2.27\,\mathrm{GeV}/c^2$ for the $\Lambda\Lambda$ (left) 
and $\Xi^-p$ (right) channels, respectively, at $E_\mathrm{cm}=2.8$ GeV. 
The width of the $H$ dibaryon is assumed to be $1$ MeV, whereas 
the phase angle $\phi$ 
is tested for $0$ (thick) and $\pi$ (thin). The shaded areas indicate 
the cases without the $H$ dibaryon. 
The light and heavy shared areas indicate the cases without 
and only with the $H$ dibaryon.
We observed that the $H$-dibaryon production rates 
are larger for the $\Xi^-p$ channel 
by a factor of two, than that for the $\Lambda\Lambda$ channel, 
and vice versa for the total background contributions, 
as shown in Fig.~\ref{fig:fig9}. 
The signal-to-background ratio is approximately $0.3$ 
for the $\Lambda\Lambda$ channel, 
whereas the larger value of $1.6$ is for the $\Xi^-p$ channel. 
Therefore, the $\Xi^-p$ channel enables us to search for 
the $H$ dibaryon significantly easier than the $\Lambda\Lambda$ channel. 
The production cross-sections for $H(2250)\to\Lambda\Lambda$ and 
$H(2270)\to\Xi^-p$ are approximately 40 nb and 38 nb, respectively.
Significant changes are obtained by the different phase factors, 
clearly shown in the $\Xi^-p$ channel, 
owing to the smaller interference with the background processes. 
Furthermore, we note that the channel opening effects 
from the final-state interactions were small, 
resulting in cusp-like structures being hardly observed. 

\begin{figure}[h]
\centering
\stackinset{r}{.5cm}{t}{0.5cm}{(a)}{
\includegraphics[width=8cm]{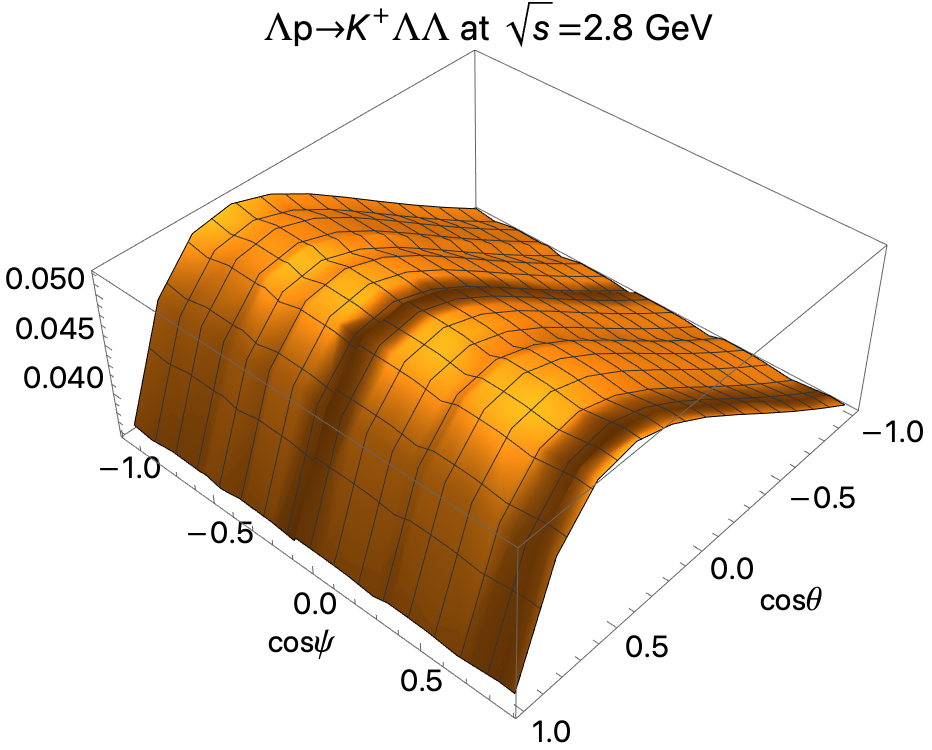}
}
\stackinset{r}{.5cm}{t}{0.5cm}{(b)}{
\includegraphics[width=8cm]{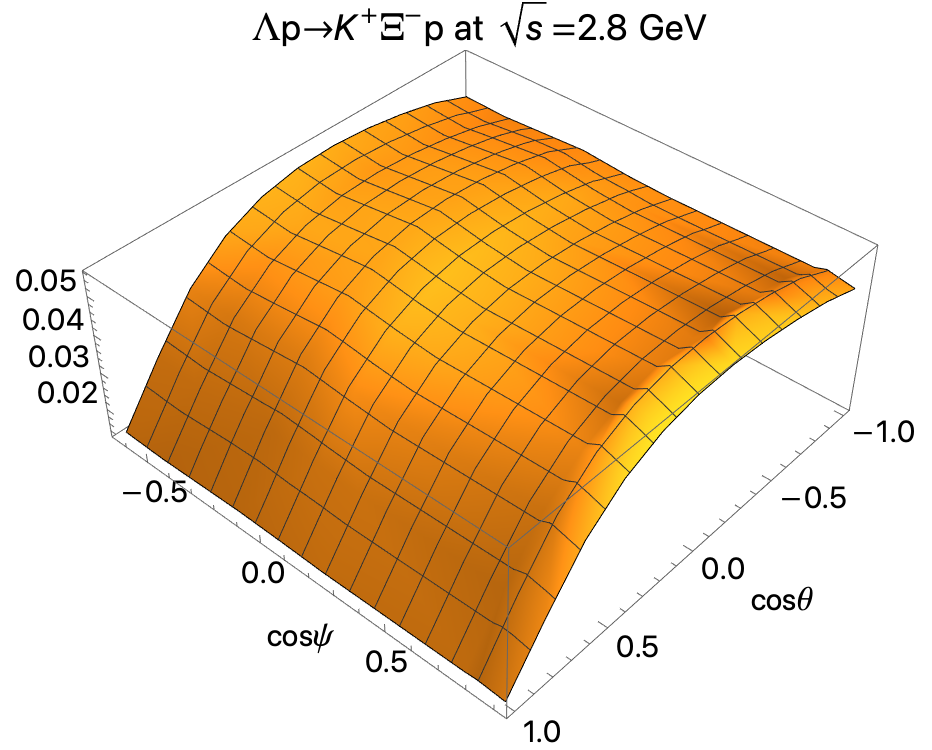}
}
\caption{(Color online) Double differential cross-section 
as a function of $\cos\theta_\mathrm{cm}$ 
and $\cos\psi_\mathrm{rest}$ for (a) the $\Lambda\Lambda$ 
and (b) $\Xi^-p$ (right) channels. See the text for details.}  
\label{fig:fig10}
\end{figure}

Finally, considering the decay-angle distribution of 
the $H$-dibaryon, 
the decay angle distribution of the $S$-wave $H$-dibaryon 
is isotropic at the rest frame of the $H$-dibaryon. 
We define the double differential cross-section as 
$d^2\sigma/d\cos\theta_\mathrm{cm}d\cos\psi_\mathrm{rest}$,
where the angle $\theta_\mathrm{cm}$ denotes that of the outgoing $K^+$ 
in the cm frame. The angle $\psi_\mathrm{rest}$ is defined by
\begin{equation}
\label{eq:PSI}
\psi_\mathrm{rest}\equiv \frac{\vec{k}^{B_f}_\mathrm{rest}\cdot \vec{k}^{K^+}_\mathrm{cm}}
{|\vec{k}^{B_f}_\mathrm{rest}||\vec{k}^{K^+}_\mathrm{cm}|},
\end{equation}
where $\vec{k}^{B_f}_\mathrm{rest}$ and $\vec{k}^{K^+}_\mathrm{cm}$ 
indicate the three momenta of the one decaying baryon 
in the final state at the $H$-dibaryon rest frame 
and the outgoing $K^+$ in the cm frame, respectively. 
In Fig.~\ref{fig:fig10}, we depict the numerical results 
for the double differential cross-sections
as a function of $\cos\theta_\mathrm{cm}$ and 
$\cos\psi_\mathrm{rest}$ with the $H$-dibaryon contribution only. 
As expected, we clearly observe that the decay-angle distribution, 
i.e., the double differential cross-sections, 
are nearly flat for the various $\cos\theta_\mathrm{cm}$ values. 

\section{Summary}
In this study, we investigated the $H(I=0,J=0)$-dibaryon production 
via $\Lambda p\to\Lambda\Lambda K^+$ theoretically. 
Thus, we employed the effective Lagrangian approach 
at the tree-level Born approximation. 
We considered the mass and decay width of the dibaryon 
as the theoretical input parameters, and they were chosen by considering 
presently available theoretical and experimental results, 
such as the lattice-QCD data analyses with the flavor SU(3) 
breaking effects: 
$2M_{\Lambda}\le M_H\le (M_{\Xi^-}+M_p)$ and $\Gamma_H=(1\sim10)$ MeV. 
The critical observations made in this study are as follows:
\begin{itemize}
\item The total cross-sections for 
the $\Lambda\Lambda$ and $\Xi^-p$ channels 
are determined to be within the order of a few $\mu$b 
in the $\Lambda$ beam momentum of up to 5 GeV$/c$, 
while the production cross-section for the $H$-dibaryon is
approximately 100 nb.
Here, we determined our model parameters 
such as the cutoff masses for the form factors, 
based on the experimental data 
for the $\Xi^-p$ elastic and $\Xi^-p\to\Lambda\Lambda$ 
scattering cross-sections.
\item The total cross-sections do not change siginificantly 
with the $H$-dibaryon mass from $2.25$ GeV$/c^2$ to $2.27$ GeV$/c^2$. 
Because the $\Lambda\Lambda$ production channel involves more background
processes than the $\Xi^-p$ channel, by ignoring the channel 
via an exotic pentaquark-state, 
the $H$-dibaryon contribution appears to be relatively large 
in the $\Xi^-p$ channel. 
\item We observed that the differential cross-sections for 
the $\Lambda p\to K^+\Lambda\Lambda$ and $\Lambda p\to K^+\Xi^-p$ channels
peak at the forward $K^+$ angles in the cm frame, 
owing to the $t$-channel meson and baryon exchange processes. 
The $H$-dibaryon-pole contributions are significant near the threshold 
and depend minimally on the $K^+$ angle.
\item From the invariant mass distributions, 
the signal-to-background ratios are approximately $0.3$ and $1.6$ 
for the $\Lambda\Lambda$ and $\Xi^-p$ channels, respectively, 
owing to the smaller background contributions in the $\Xi^-p$ channel. 
Note, the $H$-dibaryon peak areas yield 40 nb and 38 nb for 
the $\Lambda\Lambda$ and $\Xi^-p$ channels, respectively.
\item We also explored the change in the interference patterns 
between the $H$-dibaryon and background amplitudes 
with the relative phase angle
for the $\Xi^-p$ channel. The channel opening effects 
from the final-state interactions were small; therefore, 
cusp-like structures were hardly observed. 
\item Lastly, we calculated the decay angular distributions of 
the $H\to\Lambda\Lambda$ and $H\to\Xi^-p$ decays in the helicity frame 
in which the quantization axis is in the opposite direction
of $K^+$ in the $H$-dibaryon rest frame. The angular distributions are flat
over the $H$-dibaryon mass region, as expected for the $S$-wave resonance.
\end{itemize}

Considering the aforementioned factors, we conclude that 
the $H$-dibaryon could be clearly identified in 
the $\Lambda p\to K^+\Lambda\Lambda$ and $\Lambda p\to K^+\Xi^-p$
reactions close to the production threshold, if it exists close
to the $\Xi^-p$ threshold.   
Further studies related to other $H$-dibaryon production reactions 
are in progress and will appear elsewhere. 
\section*{Acknowledgment}
S.i.N. is grateful for useful discussions with T.~Hyodo. 
This work was supported in part by the National Research Foundation of Korea (NRF) 
grants funded by the Korean government (No.~2018R1A5A1025563, 
~No.~2019R1A2C1005697, ~No.~2020R1A3B2079993). 
\section*{Appendix}

\end{document}